\documentstyle[psfig,aps,twocolumn,prl]{revtex}  
\begin{document}                
\wideabs{
\title{Stripe order at low temperatures in La$_{2-x}$Sr$_x$NiO$_4$ for  $\frac{1}{3} \lesssim x \lesssim \frac{1}{2}$}
\author{H. Yoshizawa, T. Kakeshita$^{*}$, and R. Kajimoto}
\address{Neutron Scattering Laboratory, Institute for Solid State Physics, University of Tokyo,\\
 Tokai, Ibaraki 319-1106, Japan}
\author{T. Tanabe, T. Katsufuji$^{**}$, and Y. Tokura}
\address{Department of Applied Physics, University of Tokyo, Tokyo 113-8656, Japan}
\date{\today}


\maketitle

\begin{abstract} 
              
Stripe order in La$_{2-x}$Sr$_{x}$NiO$_{4}$ beyond $x = \frac{1}{3}$ was studied with neutron scattering technique.  At low temperatures, all the samples exhibit hole stripe order.  Incommensurability $\epsilon$ of the stripe order is approximately linear in the hole concentration $n_h = x + 2 \delta$ up to $x=1/2$, where $\delta$ denotes the off-stoichiometry of oxygen atoms.  The charge and spin ordering temperatures exhibit maxima at $n_h = \frac{1}{3}$, and both decrease beyond $n_h > \frac{1}{3}$.  For $\frac{1}{3} \leq n_h \lesssim \frac{1}{2}$, the stripe ordering consists of the mixture of the $\epsilon = \frac{1}{3}$ stripe order and the $n_h = \frac{1}{2}$ charge/spin order.

\end{abstract}
}


The stripe order formed by linearly segregated holes in the oxygen-doped and Sr-doped La$_{2}$NiO$_{4}$ system is studied in detail by a series of works by Tranquada and coworkers \cite{TraO105,TraO125,TraO2_15,Tra97,Tra135,Tra225}.  When the spins in this compound form the ordering at low temperatures, a hole stripe separates the antiferromagnetic Ni spin order as an antiphase domain boundary.  It was suggested that such characteristic stripe order may persist for a larger hole concentration $n_{h}$ up to $n_{h} \approx \frac{1}{2}$ with keeping the linear relation between the hole concentration $n_{h}$ and the incomensurability $\epsilon$ of the stripe order, {\it i.e.} $\epsilon \sim n_h$\cite{Tra135,Tra225}.  According to the resistivity and electron diffraction studies \cite{Che93}, on the other hand, commensurate charge order is speculated for two Sr concentrations $n_h = \frac{1}{3}$ and $\frac{1}{2}$.  In fact, very recent experimental studies have established that the $n_h =\frac{1}{3}$ sample exhibits the stripe-type charge order below $T \sim 240$K, and it accompanies with anomalies in optical conductivity and Raman spectra\cite{Kat96,Lee97,raman}.

So far, the hole stripe order in the nickelate is confirmed in the O-doped La$_{2}$NiO$_{4+\delta}$ samples with $\delta = 0.105, 0.125,  \frac{2}{15}$ (Refs. \onlinecite{TraO105,TraO125,TraO2_15,Tra97}) and the Sr-doped La$_{2-x}$Sr$_{x}$NiO$_{4}$ samples with $ x =0.135,0.20,0.225$ and $x=\frac{1}{3}$ (Refs. \onlinecite{Tra135,Tra225,Lee97}) by neutron diffraction.  For a small hole concentration $n_{h} = x + 2\delta$,  the distance between hole stripes is wide enough to accomodate three or more Ni chains in between, and this situation allows for Ni chains to form antiphase antiferromagnetic spin ordering across hole stripes.  In contrast, for the samples with larger $n_{h}$ of $\frac{1}{3} \leq n_{h} \leq \frac{1}{2}$, the distance of hole stripes is small, and only one or two Ni chains are accomodated between the hole stripes provided that the hole stripes reside on the Ni sites.  Hence it would be interesting to study whether samples with $\frac{1}{3} \leq n_{h} \leq \frac{1}{2}$ can form essentially the same stripe ordering, or they exhibit qualitatively different charge ordering.  The information on spin/charge ordering for $\frac{1}{3} \leq n_{h} \leq \frac{1}{2}$ would be also very useful to understand the behavior of the resistivity.  In order to elucidate the effects of the higher hole doping to the hole stripe order, we have carried out a neutron diffraction study on La$_{2-x}$Sr$_{x}$NiO$_{4}$ samples with the Sr concentration $x$ for $0.289 \lesssim x \lesssim 0.5$.  The preliminary results have been reported elsewhere \cite{yoshi}.

Some of the important findings in the present study are that the incommensurability $\epsilon$ in the Sr-doped nickelate is approximately linear in $n_{h}$ up to $n_{h} \approx \frac{1}{2}$, in sharp contrast with the La$_{2-x}$Sr$_{x}$CuO$_{4}$ system \cite{yam97,tra97b}.  A careful examination of the $n_{h}$ dependence of $\epsilon$ further revealed that there is a systematic deviation from an  $\epsilon \sim n_{h}$ law around $n_{h}=\frac{1}{3}$, and that such deviation strongly influences transport properties\cite{Kat99}.  We also observed that the charge ordering temperature $T_{\rm CO}$ and the spin ordering temperature $T_{\rm N}$ exhibits maxima at $n_{h}=\frac{1}{3}$, and they decrease beyond $n_{h}=\frac{1}{3}$.  In addition, the stripe order at low temperatures is of two-dimensional (2D) character, and it consists of a mixture of the $n_{h}=\frac{1}{3}$--type stripe order and the $n_{h}=\frac{1}{2}$--type charge order within the 2D NiO$_2$ planes for $\frac{1}{3} \leq n_{h} \leq \frac{1}{2}$.


Single crystal samples studied in the present study were cut from the same crystals used in the previous measurements of optical and Raman spectra as well as transport properties.  They were grown by the floating zone method, and the oxygen off-stoichiometry as well as the hole concetration $n_{h} =x + 2\delta$ were characterized in detail as previously reported  \cite{Kat96,raman,Kat99}.  All the samples are denoted by the calibrated hole concentration $n_h$ throughout this report.

The neutron scattering experiments were performed using triple axis spectrometers HQR and GPTAS installed at the JRR-3M reactor in JAERI, Tokai, Japan.  To optimize the visibility of weak signal from the charge ordering, we chose a combination of horizontal collimators of open-Sample-40$^{\prime}$-Analyzer-open (from monochromator to detector) for the HQR sperctrometer which is installed at the thermal guide tube with a fixed incident neutron momentum of 2.57{\AA}$^{-1}$.  The crystals were mounted in an Al can filled with He gas.  Following the preceding works, we denote the reciprocal space by the orthorhombic notation, and all the measurements were performed on the $ (h0l) $ scattering plane.


\begin{figure}
\centering \leavevmode
\psfig{file=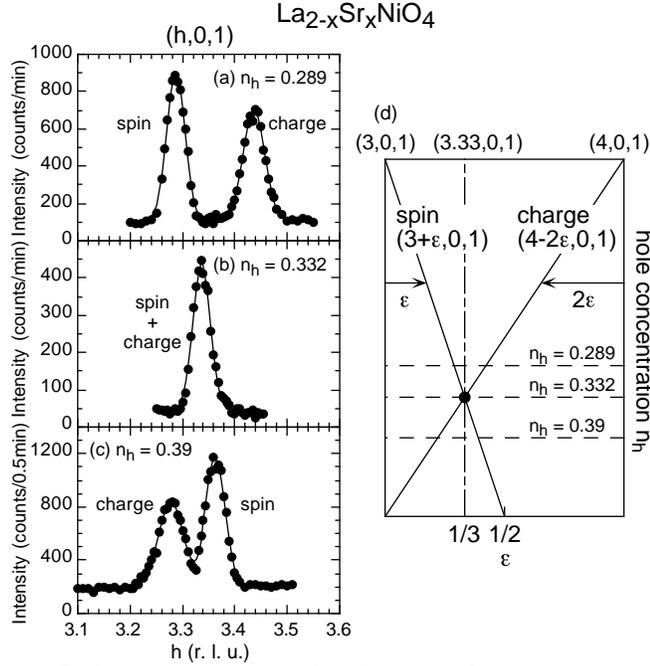,width=\hsize}
\caption{In-plane scan profiles for the spin and charge superlattice peaks observed along $(h01)$ for (a) $n_h = 0.289$, (b) $n_h = 0.332$, and (c) $n_h = 0.39$ samples. }
\label{h-profile}
\end{figure}

In order to characterize the charge and spin ordering in the highly Sr-doped nickelate samples, we first studied the ordering in the NiO$_2$ planes.  We found that the samples with $0.289 \lesssim n_h \lesssim 0.5$ show very similar superlattice reflections of the stripe order with those observed in the Sr-doped and O-doped samples with smaller $n_{h}$.  Figure \ref{h-profile} shows the profiles of the charge and spin superlattice peaks observed along the $(h01)$ line, on which the superlattice reflections of the spin order were observed at $(2n+1\pm \epsilon, 0, 1)$, while those of the charge order at $(2n\pm 2\epsilon, 0, 1)$ with $n$ integer, respectively.  Note that, due to the $\epsilon \sim n_h$ law, an increase of $n_h$ switches the relative positions of the spin and charge superlattice peaks as schematically shown in Fig. \ref{h-profile}(d).  At $n_{h}=\frac{1}{3}$, the superlattice peak of the spin order exactly coincides with that of the charge order as seen in Fig. \ref{h-profile}(b), which strongly enhances the stability of the $n_{h}=\frac{1}{3}$ stripe order and gives rise to a distinct anomaly in the resistivity\cite{Che93}.

\begin{figure}
\centering \leavevmode
\psfig{file=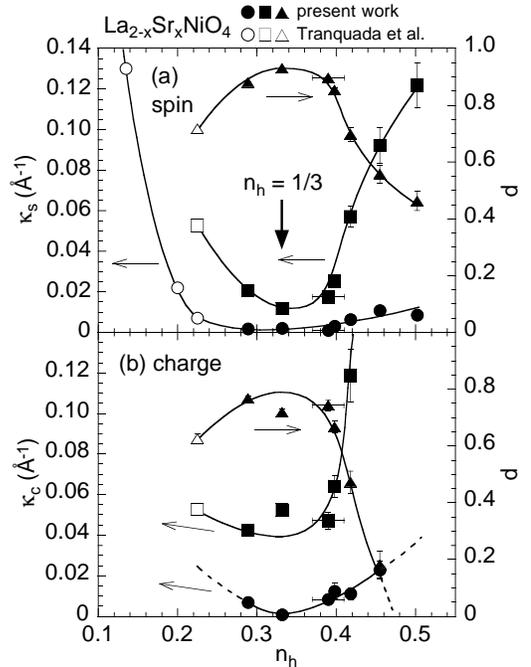,width=0.8\hsize}
\caption{Inverse correlation length of the spin order ${\kappa}_{s}$ (upper panel) and of the charge order ${\kappa}_{c}$ (lower panel) at low temperatures.  Triangles denote $p$ defined in Eq. (1).  Filled symbols are by the present work, while open symbols from Refs. \protect\onlinecite{Tra135,Tra225}.  $\kappa_c$ of $n_h=\frac{1}{3}$ is estimated from the data at 170K.}
\label{x-dep}
\end{figure}

  Reasonably sharp peaks observed in the present samples indicate that the well-developed stripe order is established within the NiO$_2$ planes up to $n_h \lesssim 0.5$.  We subsequently examined the stacking of the stripe order perpendicular to the NiO$_2$ planes, by observing the profiles along the $l$ direction (not shown).  We found that the scattering profiles are centered at $l=$ interger with stronger intensity at $l=$ odd, being similar to the results observed in the less Sr-doped samples\cite{Tra135,Tra225}.  Consequently, the inverse correlation length $\kappa$ of the stripe order along the stacking direction was evaluated by fitting to the formula suggessted by Tranquada {\it et al.} \cite{Tra225},
\begin{equation}
I(l) \sim \frac{1-p^2}{1+p^2-2 p \cos\pi l}
\end{equation}
where $p = e^{-c/2\xi_{l}} = e^{-\frac{c}{2}\kappa_{l}}$.  For $\xi_{l}/c \gg 1$, it converges to a conventional Lorentzian form.  The fact that the $l$ dependence is well described by Eq. (1) means that the correlation of the stacking of the hole stripes decays exponentially but they are resistered on the lattice at low temperatures even for $\frac{1}{3} \leq n_h \leq \frac{1}{2}$.  The profiles observed by $h$ scans are also fitted to Lorentzian, and all the results are summarized in Fig. \ref{x-dep}.  The circle symbols denote the in-plane ${\kappa}$, while squares denote ${\kappa}$ perpendicular to the NiO$_2$ planes which is evaluated from $p$.

As a function of $n_h$, we can identify three regions for the stripe order.  For $n_h \lesssim \frac{1}{4}$, the correlation length is short for both within and perpendicular to the NiO$_2$ planes, and the stripe order is essentially 3 dimensional (3D) short range order (SRO).  For $n_h \gtrsim 0.4$, on the other hand, the stripe order is well-developed within the NiO$_2$ planes, but is less correlated between the NiO$_2$ planes, being quasi-2D long range order (LRO).  Near $n_h \sim \frac{1}{3}$, $\kappa$ shows a minimum, and the stripe order is quasi-3D LRO, demonstrating the stability of the stripe order at $n_h \sim 1/3$.

\begin{figure}
\centering \leavevmode
\psfig{file=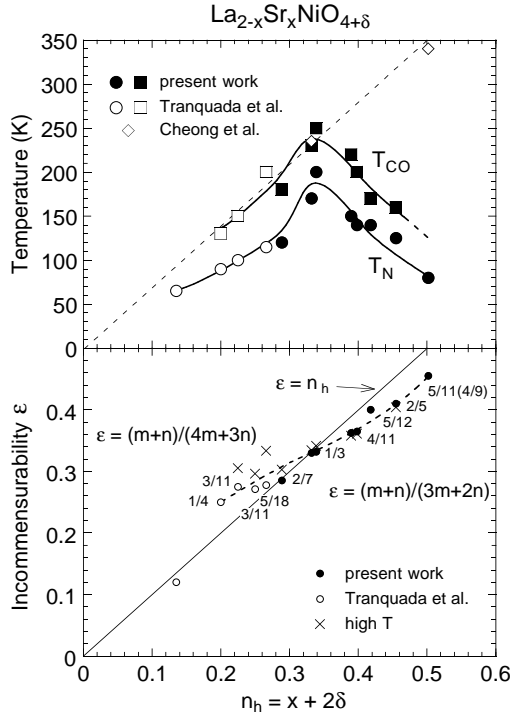,width=0.8\hsize}
\caption{Upper panel: $T_{\rm CO}$ (square and diamond symbols) and $T_{\rm N}$ (circles).  Lower panel: Hole concentration $n_h$ dependence of the incommensurability $\epsilon$ for $n_h \protect\lesssim \frac{1}{2}$ determined at the lowest temperature studied.  Cross symbols indicate the high temperature initial values of $\epsilon$. ($\Diamond$ is taken from Ref. [7], and $\bigcirc, \Box$ from Refs. [1-6].)}
 \label{n_vs_Tc}
\end{figure}

The stability of the $\epsilon=\frac{1}{3}$ stripe order is also evident in the $n_h$ dependence of the charge and spin ordering temperatures $T_{\rm CO}$ and $T_{\rm N}$, and they are summarized in the upper panel of Fig. \ref{n_vs_Tc}.  In earlier works\cite{Tra135,Tra225,Che93}, $T_{\rm CO}$ and $T_{\rm N}$ were speculated to increase linearly in $n_h$ as indicated by a dashed line for $T_{\rm CO}$.  We found, however, that they peak at $n_h =\frac{1}{3}$.  We confirmed that the transition temperatures determined in the present work are in excellent accord with the anomalies of the temperature dependence in the resistivity along the $c$ axis ({\bf E} $\parallel c$) \cite{Kat96}.  As pointed out earlier \cite{TraO125,Tra135,Tra225}, $T_{\rm CO}$ and $T_{\rm N}$ are different for all the samples we studied with $0.289 \lesssim n_h \lesssim \frac{1}{2}$, indicating that the hole stripe order is established first at $T_{\rm CO}$, and then the antiphase spin order is formed at the lower temperature $T_{\rm N}$ for $n_h \lesssim \frac{1}{2}$.

Now, we examine the incommensurability $\epsilon$ of the stripe order for $0.289 \lesssim n_h \lesssim \frac{1}{2}$ in detail.  Since $\epsilon$ is weakly temperature-dependent, $\epsilon$ of the low temperature limit is plotted at the bottom panel of Fig. \ref{n_vs_Tc} \cite{shift}.  We confirmed that $\epsilon$ is approximately linear in $n_h$ up to the limit of the stripe order, $n_h \approx \frac{1}{2}$.  This is in strong contrast to the doped cuprate superconductor La$_{2-x}$Sr$_{x}$CuO$_{4}$ and to the O-doped nickelate.  For instance, $\epsilon$ saturates at $\epsilon \sim \frac{1}{8}$ beyond the optimum doping in the cuprate\cite{yam97,tra97b}.  In the Sr-doped nickelate, on the other hand, the formation of the $\frac{1}{3}$ stripe order is clearly stable, and the region of the stripe order extends to the higher hole concentration.

\begin{figure}
\centering \leavevmode
\psfig{file=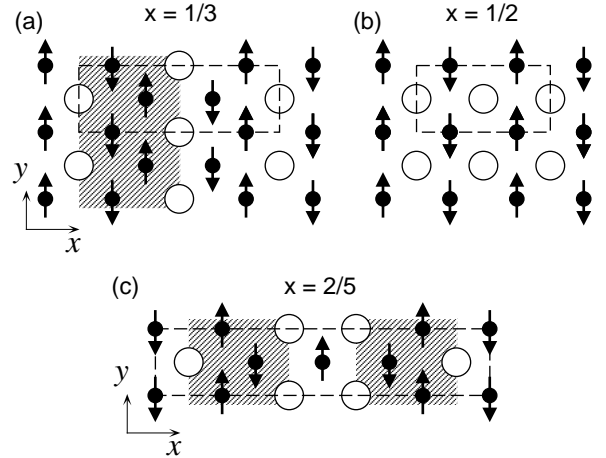,width=0.9\hsize}
\caption{Models of stripe order for $n_h=1/3$ (a), $1/2$ (b), and $n_h=2/5$ (c), respectively.  Dashed lines indicate the magnetic unit cell, while the shaded area denotes the $\frac{3}{2}a$-width unit of the $\epsilon = \frac{1}{3}$ stripe order.}
\label{CO_model}
\end{figure}

  The analysis of $\epsilon$ further provides important information on the structure of the stripe order.   For the O-doped La$_2$NiO$_{4+{\delta}}$ samples, $\epsilon$ is often locked at a rational value given by $\epsilon = (m+n)/(4m+3n)$ \cite{TraO105,TraO125,TraO2_15} because the interstitial oxygen ordering stabilizes the commensurate $\epsilon=\frac{1}{4}$ and $\frac{1}{3}$ stripe orders and introduces the competition between them.  In the present study, we found that the low temperature limit of $\epsilon$ can be expressed by the same relation for $n_h <\frac{1}{3}$, but it changes to
\begin{equation}
\epsilon = (n+m^{\prime})/(3n+2m^{\prime}) \ {\rm for}\   n_h \ge\frac{1}{3},
\end{equation}
as tabulated in Table \ref{epsilontable}.  To explain the meaning of this formula, the model of the $\epsilon=\frac{1}{3}$ stripe order and the $n_{h}=\frac{1}{2}$ charge order are depicted in Fig. \ref{CO_model}(a) and (b)\cite{comTdep}.  The fact that $\epsilon$ is given by Eq. (2) for $\frac{1}{3} \leq n_h \leq \frac{1}{2}$ can be interpreted that the stripe order in this range of hole concentration consists of a combination of the $\frac{3}{2}a$-width unit of the $\epsilon=\frac{1}{3}$ stripe order and the $a$-width unit of the $n_{h}=\frac{1}{2}$ charge order, separated by discommensuration.  An example observed in the $n_h=0.425$ sample is depicted in Fig. \ref{CO_model}(c), in which shaded regions correspond to the $\frac{3}{2}a$-width unit.  $m$, $n$, and $m^{\prime}$ in Eq. (2) and Table \ref{epsilontable} give the numbers of the $2a$-, $\frac{3}{2}a$-, and $a$-width units in the long-period commensurate unit cell.  For $\frac{1}{3} \leq n_h \leq \frac{1}{2}$, an increase of $m^{\prime}$ relative to $n$ in Table \ref{epsilontable} indicates that the fraction of the $n_{h}=\frac{1}{2}$ charge order progressively increases in the stripe ordering pattern.  A similar discommensuration pattern of stripe-type ordering was recently reported in a heavily-doped insulating manganite system La$_{1-x}$Ca$_{x}$MnO$_{3}$ for $x > \frac{1}{2}$ \cite{mori}.

The high density of hole concentration drastically influences the two ordering temperatures, $T_{\rm CO}$ and $T_{\rm N}$ for $n_h > \frac{1}{3}$.  (1) Up to $n_h = \frac{1}{3}$, the spin order is essentially an antiferromagnetic order which is separated by an antiphase domain boundary of hole stripes.  For the $n_h = \frac{1}{2}$ ordering pattern, however, the spin order is actually a 2D checkerboard pattern as depicted in Fig. \ref{CO_model}(b), and all the nearest neighbor sites of the spins are occupied by holes.  The existence of holes reduces the effective exchange interactions between spins, and lowers $T_{\rm N}$, being consistent with studies of spin dynamics in O-doped nickelates \cite{Nak,TraO97}.  (2) As seen in the right column of Table I, the discommensuration of the stripe order progressively increases the fraction of the $\epsilon = \frac{1}{2}$ pattern for $\frac{1}{3} \leq n_h \leq \frac{1}{2}$.  In the matrix of the $\epsilon =\frac{1}{3}$ stripe order, intervening $n_h =\frac{1}{2}$ patterns strongly disturb the spin/charge correlation within and between the NiO$_2$ planes as manifested by the $n_h$ depencence of $\kappa$ in Fig. \ref{n_vs_Tc}.  These effects in turn cause the suppression of $T_{\rm CO}$ and $T_{\rm N}$ for $n_h > \frac{1}{3}$.

\begin{minipage}{0.85\hsize}
\centering
\begin{table}
\caption{Incommensurability $\epsilon$ observed in the present work.  See the text for details.}
\label{epsilontable}
\begin{tabular}{cdcc}
$n_h$ & $\epsilon^{\rm obs}$ & $\frac{m+n}{4m+3n}$ & $\frac{n+m^{\prime}}{3n+2m^{\prime}}$ \\ \hline
0.289 & 0.285 & (1,1) &      \\
1/3   & 0.332 &       & (1,0)   \\
0.398 & 0.365 &       &  (3,1)   \\
0.425 & 0.398 &       &  (1,1)    \\
0.462 & 0.410 &       &  (2,3)   \\
1/2 & $\sim$0.455 &   &  (1,3) or (1,4)
\end{tabular}
\end{table}
\end{minipage}

Finally, we point out that $\epsilon$ was slightly shifted towards $\epsilon = \frac{1}{3}$ for both sides of $n_h =\frac{1}{3}$ as indicated by a dashed curve as shown in Fig. \ref{n_vs_Tc}.  One can see that farther the distance of $n_h$ from $n_h =\frac{1}{3}$, larger the deviation of $\epsilon$ from the $\epsilon \sim n_h$ law.  This behavior has an interesting implication in the transport properties \cite{Kat99}.  In the stripe model, the hole density in a stripe $n_{\rm st}$ is always $n_{\rm st}=1$ for all $\epsilon$ when the $\epsilon \sim n_h$ law holds.  Here, the hole density in a stripe $n_{\rm st}$ is defined as $n_{\rm st} \equiv$ \{number of holes/Ni site\}/\{number of domain walls (DW)/Ni site\} $= n_h / \epsilon$.  Because one hole exists per each Ni sites, hole stripes are half filled and they are Mott insulator-like.  The deviation of $\epsilon$ from the $\epsilon \sim n_h$ law indicates that the hole density deviates from $1$ for both sides of $n_h = \frac{1}{3}$.  For $n_h < \frac{1}{3}$, $n_{\rm st} \lesssim 1$, and the carriers are expected to be electron-like, while for $n_h >\frac{1}{3}$, $n_{\rm st} \gtrsim 1$, and the carriers are hole-like.  This consideration is fully consistent with the change of the sign of the Hall coefficient $R_{\rm H}$ at $n_h = \frac{1}{3}$ \cite{Kat99}.

In summary, we have presented that the region of the stripe order extends up to $n_h \sim \frac{1}{2}$  at low temperatures $\sim$10 K in La$_{2-x}$Sr$_{x}$NiO$_4$.  Incommensurability $\epsilon$ shows the $\epsilon \sim n_h$ law with systematic deviation around $n_h = \frac{1}{3}$, which controls the nature of carriers in hole stripes.  The stripe order consists of combination of the $\epsilon = \frac{1}{3}$ stripe order and the $n_h = \frac{1}{2}$ charge order, and $\epsilon$ is given by $(m+n)/(4m+3n)$ for $n_h < \frac{1}{3}$ or by $(n+m^{\prime})/(3n+2m^{\prime})$ for $n_h > \frac{1}{3}$.  The $n_h = \frac{1}{3}$ stripe order is stabilized by the coincidence of the periodicities of the charge, and spin order, and as a result, it forms quasi-3D LRO, while it is 3D SRO for $n_h < \frac{1}{4}$, and quasi-2D LRO for $0.4 \lesssim n_h \lesssim \frac{1}{2}$.

We thank J. M. Tranquada for valuable discussions.  This work was supported by a Grant-In-Aid for Scientific Research from the Ministry of Education, Science and Culture, Japan and by the New Energy and Industrial Technology Development Organization (NEDO) of Japan.


\begin{references}
\bibitem{shift} $\epsilon$ is temperature-dependent, and the initial values of $\epsilon$ at high temperature are indicated cross symbols in Fig. \ref{n_vs_Tc}.  The present samples with $n_h \gtrsim 0.289$ exhibited a shift of $0.01 \sim 0.02 a^{*}$, but it was much smaller than those of the samples reported in earlier works \cite{TraO125,TraO2_15,Tra225}.%
\bibitem{comTdep} The models of the charge/spin ordered pattern are schematically depicted for the Ni-centered stripe case.  When $T$ decreases, however, the density of O-cetered hole stripes decreases, and the Ni-centered hole stripes dominate at low temperatures, as argued in Ref. \onlinecite{TraO2_15}.  Our results are consistent with this view.
\end{references}
\end{document}